\documentclass{article}

\usepackage{PRIMEarxiv}

\usepackage[utf8]{inputenc} 
\usepackage[T1]{fontenc}    
\usepackage{hyperref}       
\usepackage{url}            
\usepackage{booktabs}       
\usepackage{amsfonts}       
\usepackage{nicefrac}       
\usepackage{microtype}      
\usepackage{lipsum}
\usepackage{fancyhdr}       
\usepackage{graphicx}       
\graphicspath{{image/}}     
\RequirePackage[fleqn]{amsmath}

\usepackage{subfig}
\usepackage{subfiles} 

\usepackage{multirow}

\title{ADLight: A Universal Approach of Traffic Signal Control with Augmented Data Using Reinforcement Learning}

\author{%
  \textbf{Maonan Wang}\\
  School of Science and Engineering, The Chinese University of Hong Kong, Shenzhen \\
  Shanghai AI Laboratory, Shanghai, China \\
  Email: \href{mailto:maonanwang@link.cuhk.edu.cn}{maonanwang@link.cuhk.edu.cn} \\
  \And
  \textbf{Yutong Xu}\\
  School of Data Science, The Chinese University of Hong Kong, Shenzhen \\
  Email: \href{mailto:yutongxu@link.cuhk.edu.cn}{yutongxu@link.cuhk.edu.cn} \\
  \And
  \textbf{Xi Xiong}\\
  Shenzhen Research Institute of Big Data \\
  Email: \href{mailto:xiongxi@cuhk.edu.cn}{xiongxi@cuhk.edu.cn} \\
  \And
  \textbf{Yuheng Kan} \\
  SenseTime Group Limited, Shanghai, China \\
  Shanghai AI Laboratory, Shanghai, China \\
  Email: \href{mailto:kanyuheng@sensetime.com}{kanyuheng@sensetime.com} \\
  \And
  \textbf{Chengcheng Xu} \\
  SenseTime Group Limited, Shanghai, China \\
  Email: \href{mailto:xuchengcheng@sensetime.com}{xuchengcheng@sensetime.com} \\
  \And
  \textbf{Man-On Pun} \\
  School of Science and Engineering, The Chinese University of Hong Kong, Shenzhen \\
  Shenzhen Research Institute of Big Data \\
  Email: \href{mailto:simonpun@cuhk.edu.cn}{simonpun@cuhk.edu.cn} {\em (Corresponding Author)}
}

\begin{document}
\maketitle

\section{Abstract}

Traffic signal control has the potential to reduce congestion in dynamic networks. Recent studies show that traffic signal control with reinforcement learning (RL) methods can significantly reduce the average waiting time. However, a shortcoming of existing methods is that they require model retraining for new intersections with different structures. In this paper, we propose a novel reinforcement learning approach with augmented data (ADLight) to train a universal model for intersections with different structures. We propose a new agent design incorporating features on movements and actions with \textit{set current phase duration} to allow the generalized model to have the same structure for different intersections. A new data augmentation method named \textit{movement shuffle} is developed to improve the generalization performance. We also test the universal model with new intersections in Simulation of Urban MObility (SUMO). The results show that the performance of our approach is close to the models trained in a single environment directly (only a $5\%$ loss of average waiting time), and we can reduce more than $80\%$ of training time, which saves a lot of computational resources in scalable operations of traffic lights. 

\noindent\textit{Keywords}: traffic signal control, model generalization, reinforcement learning, data augmentation

\section{Introduction}

Traffic congestion has been one of the major challenges in many metropolitan areas. 
Congestion can not only increase travel time in networks, but also result in environmental problems, e.g., fuel waste and pollutant emissions \cite{kweku2018greenhouse}.
One way to address this issue is to control traffic lights dynamically. 
Nowadays, most traffic lights are controlled with pre-defined fixed-time plans \cite{miller1963settings}. 
Conventional traffic signal control methods, such as the Webster model \cite{koonce2008traffic} and Self-Organizing Traffic Light Control (SOTL) \cite{gershenson2004self}, are also used in this domain.
However, these methods rely heavily on expert knowledge and certain assumptions about the traffic model, e.g., uniform traffic flow distribution \cite{roess2004traffic}, which is not realistic in real applications \cite{shabestaryAdaptiveTrafficSignal2022}.

In order to recognize traffic patterns from real traffic data, some reinforcement learning methods are proposed to solve this problem \cite{xu2013study, mannion2016experimental, van2016coordinated, weiIntelliLightReinforcementLearning2018, weiPressLightLearningMax2019, weiCoLightLearningNetworklevel2019, zhengLearningPhaseCompetition2019, liangDeepReinforcementLearning2019, chenThousandLightsDecentralized2020, ma2021adaptive, shabestaryAdaptiveTrafficSignal2022, mousavi2017traffic, aslani2017adaptive, aslaniTrafficSignalOptimization2018, xiongLearningTrafficSignal2019, yangCooperativeTrafficSignal2019, oroojlooy2020attendlight}. 
\cite{weiIntelliLightReinforcementLearning2018} proposed a deep Q-learning algorithm with the phase gate to control signals on a large-scale network. 
\cite{xiongLearningTrafficSignal2019} used the Advantage Actor Critic (A2C) method to train the agent and leverage demonstrations collected from classic methods to accelerate the learning process. 
By directly interacting with the environment, an RL agent can learn effectively how to adapt to changes in traffic conditions with real experience.

In spite of the significant improvement with RL methods in this domain, 
the major limitation is that the model needs to be redesigned and trained from scratch when faced with intersections of different structures, 
e.g., different approaching roads, different lanes, and different phases. 
It will take much time if we choose to learn the optimal policy for each intersection in large-scale networks.
There exist some works to design a universal model for different junctions \cite{zhengLearningPhaseCompetition2019, chenThousandLightsDecentralized2020, oroojlooy2020attendlight}; 
however, all these methods utilize the \textit{choose next phase} as the action design, which would lead to safety concerns since this method breaks the traditional traffic cycle by randomly choosing phases.

To address the problem above, we present the ADLight, a novel approach that can be used to train a universal reinforcement learning model with augmented data for the traffic signal control problem. 
To address the intersections with the different structures, we perform feature extraction by movement, instead of lanes. 
The features of each movement include not only traffic flows but also the movement itself. 
To ensure safety and generalization with different phase structures, \textit{set current phase duration} is chosen as the action in our ADLight. 
With this design, the agent can select actions from a list of pre-defined time periods to establish the duration for the current phase. 
A new data augmentation method \textit{movement shuffle} is developed in the ADLight, and the results show that it can improve the performance significantly. 
Experiments are conducted on intersections with different approaching roads, different lanes, and different phases in Simulation of Urban MObility (SUMO) \cite{krajzewicz2002sumo}. 
The experiments show that the resulting model from ADLight can achieve satisfactory performance in untrained intersections with different structures. 

Our contributions can be summarized as below:

\begin{itemize}
    \item We propose an agent design that allows the model to have the same structure at different intersections. 
    This paper is, to the best of our knowledge, the first effort that utilizes the \textit{set current phase duration} to create universal models for the traffic signal control problem.
    \item A new data augmentation method for traffic signal control is proposed in our approach. 
    The performance can be further improved by incorporating the data augmentation into the agent design.
    \item We demonstrate the generalization performance of our ADLight at $11$ intersections with different structures. The results show that when compared with the model trained on a single environment from scratch (which takes around $6 \times 10^{6}$ steps), the loss of the average waiting time in our generalization model can be reduced to $5\%$ after only $1 \times 10^{6}$ steps by retraining the universal model. When a network has 1000 intersections, we can save more than $5 \times 10^{9}$ interactions with SUMO, thereby greatly reducing computing resources consumption.
\end{itemize}

The rest of the paper is organized as follows. 
We first summarize the related work on traffic signal control. 
The road and traffic signal terminology is introduced in the following section. This is followed by the \nameref{methodology} section, which formally describes the ADLight. We then show the dataset and the experiments in the SUMO platform. Finally, we provide the conclusions and future directions.

\section{Related Work} \label{related_work}

\textbf{Conventional Traffic Signal Control.} 
Several classical methods have been developed to reduce the total delay for all vehicles \cite{weiSurveyTrafficSignal2020}.  
For example, the Webster \cite{koonce2008traffic} method is one of the most widely-used methods for the single intersection in this field. 
It determines the optimum cycle length and phase split for a single intersection according to traffic volume. 
In this research, the Webster method is taken into comparison as a baseline for RL-based methods. 
Self-Organizing Traffic Light Control (SOTL) \cite{gershenson2004self} is a fully-actuated control  algorithm. 
It decides whether to keep or change the current phase based on whether the number of vehicles approaching the green signal is larger than a threshold. 
The detailed rules for SOTL can be found in \cite{weiSurveyTrafficSignal2020}. 
SCATS (Sydney Coordinated Adaptive Traffic System) \cite{lowrie1990scats} is an intelligent transportation system that is widely used at more than $50,000$ intersections in over $180$ cities in $28$ countries. 
It selects from pre-defined traffic signal plans (i.e., cycle length, phase split and offsets) according to the data derived from loop detectors or other road traffic sensors. 
However, these conventional methods are usually based on oversimplified information, assumptions about the traffic model (i.e., assuming the traffic flow is uniform during a certain period), 
or need expert knowledge to design the pre-defined signal plans. 

\textbf{RL-based Traffic Signal Control.} 
With the success of deep reinforcement learning (DRL) in different areas \cite{mnih2013playing, wang2020deep}, 
more and more research studies are trying to use DRL to solve traffic signal problems \cite{haydariDeepReinforcementLearning2020}. 
A number of works \cite{xu2013study, mannion2016experimental, van2016coordinated, aslani2017adaptive, weiIntelliLightReinforcementLearning2018, weiPressLightLearningMax2019, weiCoLightLearningNetworklevel2019, zhengLearningPhaseCompetition2019, liangDeepReinforcementLearning2019, chenThousandLightsDecentralized2020, ma2021adaptive, shabestaryAdaptiveTrafficSignal2022} use value-based methods 
while others \cite{mousavi2017traffic, aslaniTrafficSignalOptimization2018, xiongLearningTrafficSignal2019, yangCooperativeTrafficSignal2019, oroojlooy2020attendlight} use policy-based methods. 
These works vary in the action designs, 
including \textit{choose next phase} \cite{weiPressLightLearningMax2019, weiCoLightLearningNetworklevel2019, zhengLearningPhaseCompetition2019, chenThousandLightsDecentralized2020, shabestaryAdaptiveTrafficSignal2022, yangCooperativeTrafficSignal2019, oroojlooy2020attendlight}, 
\textit{keep or change} \cite{mannion2016experimental, van2016coordinated, weiIntelliLightReinforcementLearning2018}
and \textit{set current phase duration} \cite{xu2013study, aslani2017adaptive, aslaniTrafficSignalOptimization2018}. 

A few attempts at training general models for traffic signal control have been made by previous research works. 
For instance, the FRAP \cite{zhengLearningPhaseCompetition2019} method is proposed to adapt to new scenarios. 
\cite{chenThousandLightsDecentralized2020} adapt the parameter sharing method based on FRAP and show strong performance on the scale of thousands of traffic lights. 
Nevertheless, they all utilize \textit{choose next phase} as the action design, and it cannot keep the original phase structure of traffic lights. 
At each action decision step, the signal phase can be chosen from all possible phases that are combined from all non-conflicting movements rather than the original phases, which ignores the pedestrian and are against driving habits. 
AttendLight \cite{oroojlooy2020attendlight} incorporates the attention mechanism to train a universal model for the intersections with different structures and traffic flow distribution. 
Although it can maintain the phase structure, its action design is also \textit{choose next phase}, 
which may make traffic signals change in a random sequence. 
This can be an issue of concern in practice as it leads to an unsafe situation for both drivers and pedestrians \cite{liangDeepReinforcementLearning2019}. 
In sharp contrast, \textit{set current phase duration} is used in our research, which is more reasonable and efficient than \textit{choose next phase}.

\section{Preliminary Definitions} \label{Preliminary}

\begin{figure}[ht!]
    \centering
    \subfloat[4-way intersection]{
        \includegraphics[width=0.43\textwidth]{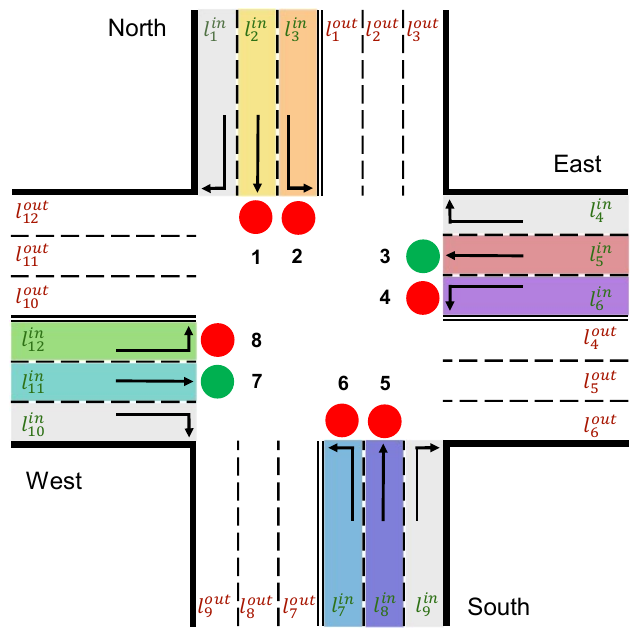}
        \label{fig_background_4way_intersection}
    }
    \subfloat[8 movements signals and the associated lanes for each signal]{
        \includegraphics[width=0.43\textwidth]{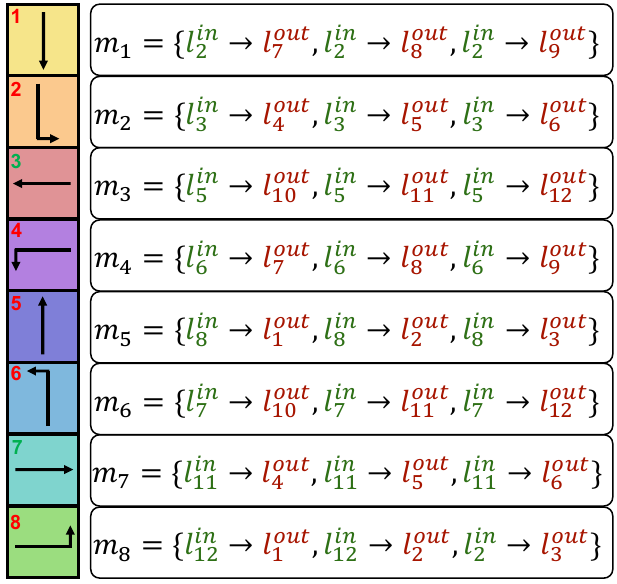}
        \label{fig_background_movement}
    }
    \quad
    \subfloat[4 phases and the associated movement signals and duration for each phase]{
        \includegraphics[width=0.43\textwidth]{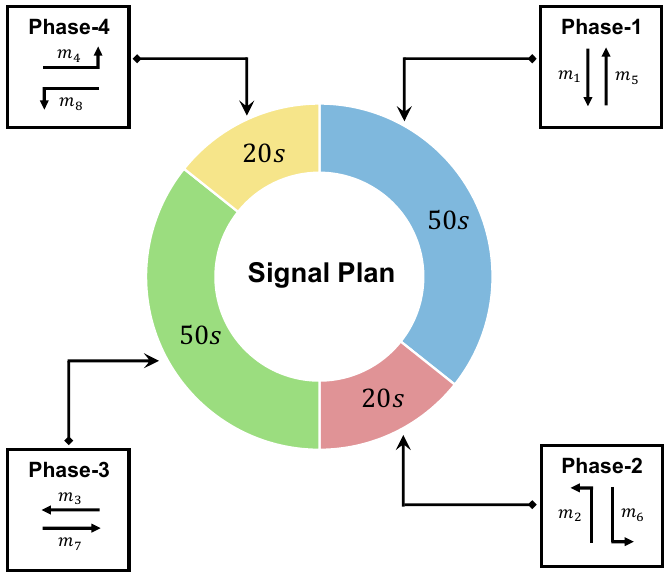}
        \label{fig_background_phase}
    }
    \caption{A standard 4-way intersection with its movement signals and phases.}
    \label{fig_background}
\end{figure}

A standard four-way intersection shown in Figure~\ref{fig_background_4way_intersection} is used as an example to illustrate the terminology. 
These concepts can be easily extended to the intersections with different structures. 

\begin{itemize}
    \item \textbf{Traffic movement: }
    A traffic movement is a connection between an incoming lane $l_{in}$ to an outgoing lane $l_{out}$, denoted as $l_{in} \rightarrow l_{out}$. 
    For the common 4-way intersection in Figure~\ref{fig_background_4way_intersection}, 
    there are a total of $12$ movements, including straight, left and right turns in four directions. 
    \item \textbf{Movement signal: }
    A movement signal is defined on the traffic movement. 
    The green signal means the corresponding movement is allowed and the red signal indicates the movement is prohibited. 
    In Figure~\ref{fig_background_4way_intersection}, the movement signal $3$ is green, indicating that the vehicle can travel from east to west at this time. 
    Although there are $12$ movements in a 4-way intersection, as the right-turn traffic can pass regardless of the signal, 
    only eight movement signals are used. 
    Figure~\ref{fig_background_movement} shows the eight movements signals and the incoming lanes and outgoing lanes associated with each signal. 
    For example, $m_{3}$ indicates that the vehicles can go from $l_{5}^{in}$ to $l_{10}^{out}$, $l_{11}^{out}$ and $l_{12}^{out}$ respectively. 
    \item \textbf{Phase: }
    A phase is a combination of movement signals. 
    Figure~\ref{fig_background_phase} shows the four phases of the 4-way intersection. 
    Each phase involves a set of movement signals. 
    For example, phase-1 involves $m_{1} = \{l_{2}^{in} \rightarrow l_{7}^{out}, l_{2}^{in} \rightarrow l_{8}^{out}, l_{2}^{in} \rightarrow l_{9}^{out}\}$ and $m_{5} = \{l_{8}^{in} \rightarrow l_{1}^{out}, l_{8}^{in} \rightarrow l_{2}^{out}, l_{8}^{in} \rightarrow l_{3}^{out}\}$. 
    It should be noted that the number of movement signals contained in different phases may vary. 
    \item \textbf{Signal plan: }
    A signal plan for an intersection is a sequence of phases and their corresponding durations. 
    Here we denote a signal plan as $\{(p_{1}, t_{1}), (p_{2}, t_{2}), \cdots, (p_{i}, t_{i}), \cdots\}$, 
    where $p_{i}$ and $t_{i}$ represent a phase and the duration of this phase, respectively. 
    Usually, the phase sequence is in a cyclic order; 
    that is, the signal plan can be denoted as $\{(p_{1}, t_{1}), (p_{2}, t_{2}), \cdots, (p_{N}, t_{N}), (p_{1}, t_{1}), (p_{2}, t_{2}), \cdots, (p_{N}, t_{N}), \cdots\}$. 
    Figure~\ref{fig_background_phase} shows a cycle-based signal plan and the duration of each phase is $50$, $20$, $50$ and $20$ as an example. 
    For the signal plan optimization task, the common methods mainly adjust each phase $p_{i}$ duration $t_{i}$ to achieve the purpose of reducing the total waiting time of  vehicles at the intersection. 
\end{itemize}

\section{Methodology} \label{methodology}

\begin{figure}[!ht]
  \centering
  \includegraphics[width=0.99\textwidth]{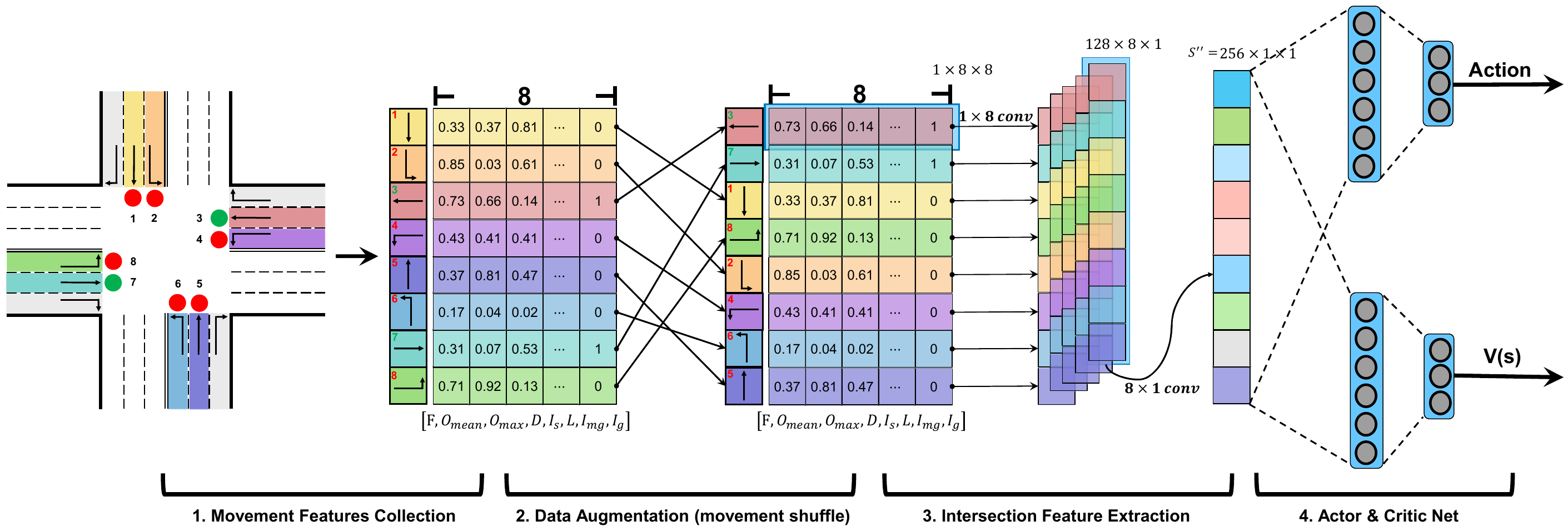}
  \caption{The overall structure of ADLight.}
  \label{fig_methodology_framework}
\end{figure}

In this section, the ADLight, as shown in Figure~\ref{fig_methodology_framework}, is proposed to train a universal model for the traffic signal control problem. 
There are four parts in the ADLight: 
(1) \textit{Movement Features Collection}. 
In order to allow the model to have the same structure at crossings with various topologies and signal schemes, a specific agent design is presented to incorporate features on movements and actions with \textit{set current phase duration}. 
(2) \textit{Data Augmentation}. 
We propose a new data augmentation technique for traffic signal control that can greatly enhance the models' performance. 
(3) \textit{Intersection Feature Extraction}. 
The generalization of the model is further improved by applying parameter sharing among all the movements and intersections. 
(4) \textit{Actor \& Critic Net}. After obtaining information from intersections, an actor net is used to predict the duration of the current phase. 
The agent design and the RL algorithm are also included in this section.

\subsection{Agent Design and Feature Collection} \label{methodology_agent_design}

\begin{figure}[ht!]
    \centering
    \subfloat[3-way intersection]{
        \includegraphics[width=0.45\textwidth]{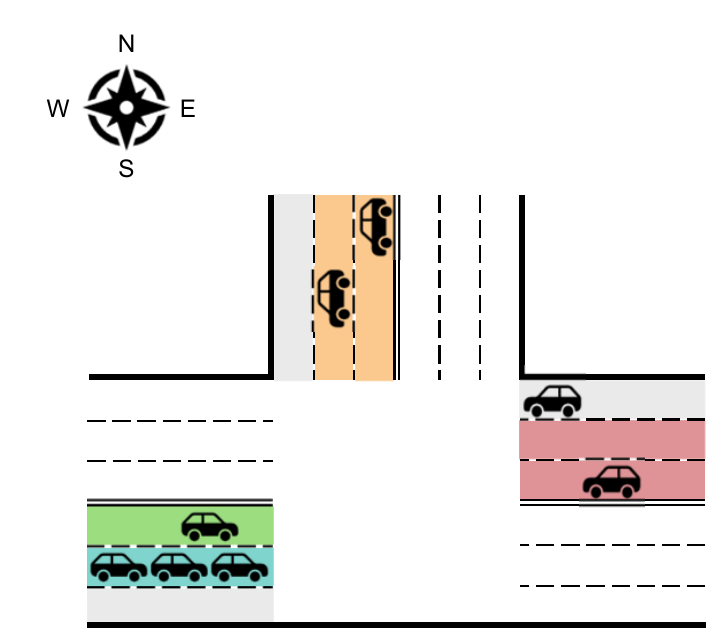}
        \label{fig_background_3_way}
    }
    \subfloat[Zero padding on the state of the 3-way intersection]{
        \includegraphics[width=0.45\textwidth]{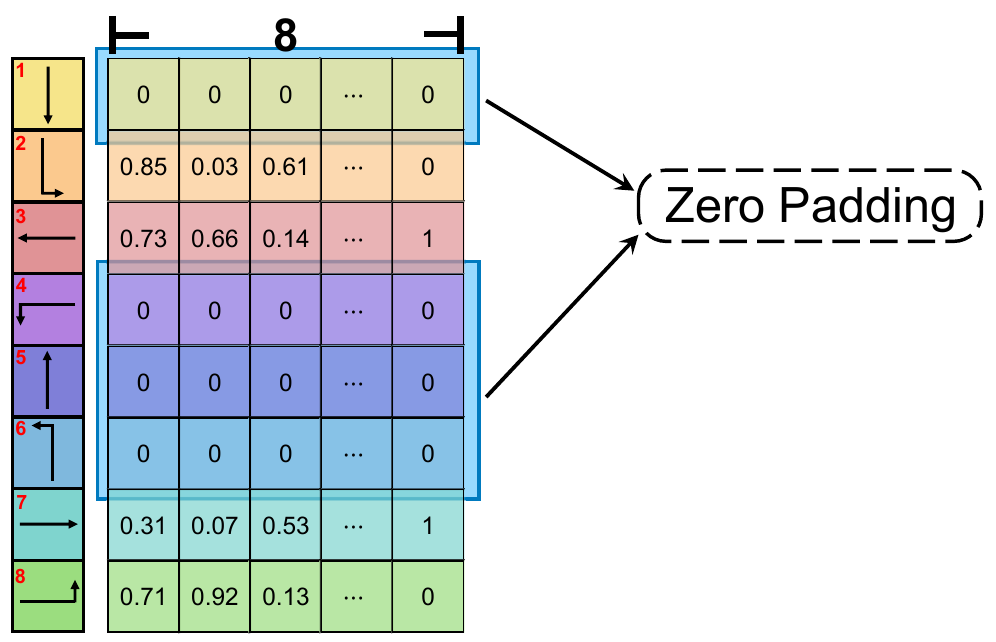}
        \label{fig_background_zero_padding}
    }
    \caption{A 3-way intersection with its state.}
    \label{fig_methodology_state}
\end{figure}

For traffic signal control, the RL agent is designed as follows:

\textbf{State}: 
Different intersections have different numbers of lanes, and if the feature is recorded in units of lanes \cite{weiIntelliLightReinforcementLearning2018, oroojlooy2020attendlight, shabestaryAdaptiveTrafficSignal2022}, the shape of the state space will be different. 
As discussed in Section~\nameref{Preliminary}, 
there are eight movement signals, no matter how many lanes a 4-way intersection has. 
More specifically, the eight movement signals are N, NL, E, EL, W, WL, S, and SL (Northbound, Northbound Left-turn, Eastbound, Eastbound Left-turn, Westbound, Westbound Left-turn, Southbound, Southbound Left-turn). 
Therefore, in this research, the state is defined as $\mathbf{S}=[\vec{m}_{1}, \vec{m}_{2}, \cdots, \vec{m}_{8}]^\mathsf{T}$, 
where $\vec{m}_{i}$ represents the information extracted by the i-th movement. 
It should be noted that when the number of movements at an intersection is less than 8 (such as a 3-way intersection), zero padding is used. 
For example, Figure~\ref{fig_background_3_way} shows a common 3-way intersection; 
only 4 movement signals are in use here, those are E, EL, W, and SL. 
As shown in Figure~\ref{fig_background_zero_padding}, zero padding is used on the rest of the 4 movement signals to keep the size of the state the same as the 4-way intersection. 

As shown in Eq.~\eqref{eq_agentDesign_1}, 
each $\vec{m}_{i}$ features three parts, \textit{traffic characteristics}, \textit{movement characteristics} and \textit{traffic signal characteristics}. 
\textit{Traffic characteristics} indicate the congestion of the movements; 
it includes the average traffic flow $F$, the maximum occupancy $O_{max}$ and the average occupancy $O_{mean}$ within $k$ seconds. 
\textit{Movement characteristics} contain information about the movement itself, 
that is the direction of the movement (whether it is going straight) $I_{s}$ and how many lanes the movement contains $L$. 
\textit{Traffic signal characteristics} include 
the duration of the movement $D$, 
whether the duration torches the min green $i_{mg}$, and whether the movement signal is green now $I_{g}$. 
The above eight features can be easily obtained, which makes this design practical for deployment.

\begin{equation} \label{eq_agentDesign_1}
\vec{m_{i}} = (
            \underbrace{Flow, Occupancy_{mean}, Occupancy_{max}}_{Traffic}, \quad
            \underbrace{Is_{straight}, Lanes}_{Movement}, \quad
            \underbrace{Duration, Is_{min\_green}, Is_{green}}_{Traffic \quad Signal}
        ).
\end{equation}

\textbf{Action}: 
A realistic and implementable action design needs to consider the safety of all traffic participants. 
Although the action design \textit{choose next phase} \cite{chenThousandLightsDecentralized2020, oroojlooy2020attendlight, shabestaryAdaptiveTrafficSignal2022} can greatly enhance the efficiency of the intersection, 
this action design will change the original phase sequence, thus reducing the safety of drivers. 
Therefore, the action design \textit{set current phase duration} \cite{xu2013study, aslani2017adaptive, aslaniTrafficSignalOptimization2018} is adopted in this experiment. 
It follows the concept of cycles, and each phase is executed sequentially (i.e., phase 1, phase 2, phase 3, phase 4, phase 1, phase 2 ...). At the beginning of each phase, the agent selects a green time duration from the set $[5s, 10s, 15s, 20s, 30s, 35s, 40s, 45s, 50s, 55s, 60s, \\ 65s, 70s]$ for this phase according to the traffic condition $\mathbf{S}$. 
\textit{Set current phase duration} only changes the duration for each phase; therefore, it has good scalability for intersections with different signal plans. 

\textbf{Reward}: 
The negative of the average queue length in each movement $q_{m_{i}}$ is used as a reward. 
Metrics such as waiting time, travel time, and delay are not used since they are hard to obtain from real-world traffic detection devices. 
We normalize the reward to limit its range and speed up the training process, represented as Eq~\eqref{eq_agentDesign_2}:

\begin{equation} \label{eq_agentDesign_2}
    r_{t} = \frac{-\sum_{i=1}^{8}{q_{\vec{m_{i}}}} - \mu}{\sigma + \epsilon}, \\
    \text{in which} \ 
    \mu=(\sum_{j=1}^{t-1}{r_{j}})/(t-1), \
    \sigma = \sqrt{\frac{1}{t-1} \sum_{j=1}^{t-1}{(r_{j} - \mu)^{2}}}.
\end{equation}

\subsection{Data Augmentation}

Recently, some research works have shown that data augmentation can improve the generalization in RL \cite{cobbe2019quantifying, lee2019network}, which means the agent can handle unseen tasks simply by training on more diversely augmented samples. 
Inspired by this, we propose new data augmentation methods called \textit{movement shuffle}, which will strongly help to improve the generalization of our method. 

Once the traffic condition $\mathbf{S}$ is obtained from an intersection, 
then we shuffle the row of $\mathbf{S}$ to get the new condition $\mathbf{S^{'}}$. 
For example, in the second part of Figure~\ref{fig_methodology_framework}, 
the raw state is $\mathbf{S}=[\vec{m}_{1}, \vec{m}_{2}, \vec{m}_{3}, \vec{m}_{4}, \vec{m}_{5},  \vec{m}_{6}, \vec{m}_{7}, \vec{m}_{8}]^\mathsf{T}$, 
and it becomes $\mathbf{S^{'}}=[\vec{m}_{3}, \vec{m}_{7}, \vec{m}_{1}, \vec{m}_{8}, \vec{m}_{2}, \vec{m}_{4}, \vec{m}_{6}, \vec{m}_{5}]^\mathsf{T}$ after movement shuffling. 

\begin{figure}[!ht]
  \centering
  \includegraphics[width=0.99\textwidth]{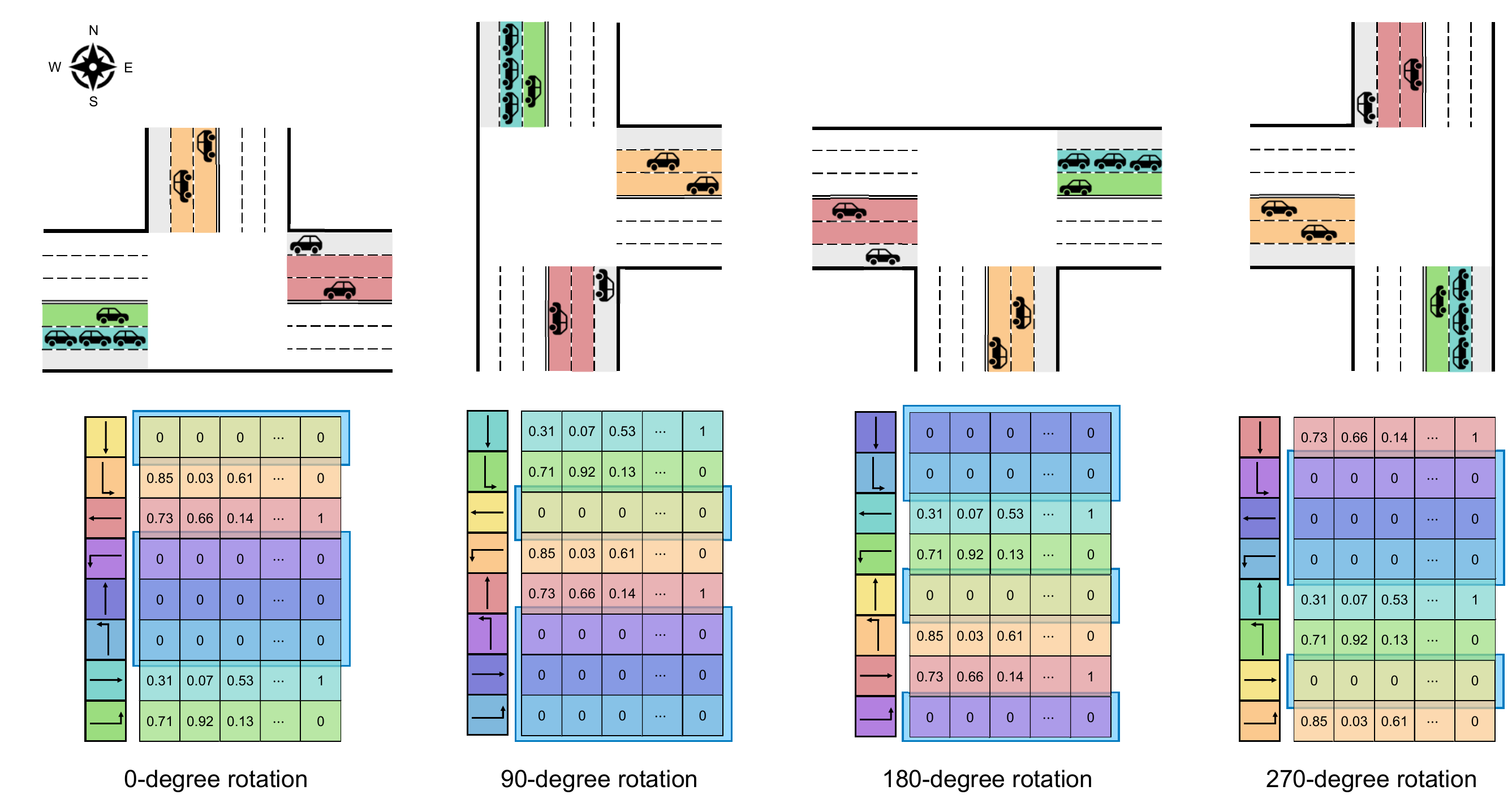}
  \caption{A 3-way intersection and the variations based on rotation. Ideally, an RL agent should output the same action for all these cases.}
  \label{fig_methodology_state_augmentation}
\end{figure}

We believe this \textit{movement shuffle} is in line with the characteristics of the intersection. 
Figure~\ref{fig_methodology_state_augmentation} shows a 3-way intersection and three variations of 90, 180, and 270 degree rotation, respectively. 
Such rotation changes the order of $\vec{m}_{i}$ in the $\mathbf{S}$ and results in a different state representation. 
For example, in the intersection without rotation, the vehicles cannot go from north to south, so $\vec{m}_{1}=\mathbf{0}$. 
In contrast, for the intersection with a 90-degree rotation, vehicles can go straight from south to north, so $\vec{m}_{1}$ is no longer $\mathbf{0}$. 
Furthermore, $\vec{m}_{1}$ in the original rotation intersection becomes $\vec{m}_{3}$ at the intersection with a 90-degree rotation. 
An RL agent that learns one case can hardly handle the other three cases. 
Nevertheless, it is reasonable to assume that the action taken by the agent should not change only after rotating the intersection. 
Therefore, \textit{movement shuffle} is used to let the model neglect the different orders of $\vec{m}_{i}$.

\subsection{Intersection Feature Extraction}

To further enhance generalization, 1D convolution layers are used to extract information for movement and intersection. 
As shown in the third part of Figure~\ref{fig_methodology_framework}, 
a 1D convolution layer with the input channel of size $1$, kernel size $1 \times 8$, and the output channel of size $128$ is used to extract information for each movement, 
which ensures this convolution layer has the generalization to the movement. 
After this convolution layer, the shape of the data is converted from $1 \times 8 \times 8$ to $128 \times 8 \times 1$. 
After that, a 1D convolution layer with an input channel of size $128$, kernel size $8 \times 1$, and an output channel of size $256$ is used to extract information for the intersections. 
Finally, the \textit{intersection feature extraction} part outputs a vector $\mathbf{S^{''}}$ of length $256$; 
this vector contains the information of an intersection.

\subsection{Reinforcement Learning Framework}

Proximal Policy Optimization (PPO) \cite{schulman2017proximal} is used to solve this RL problem. 
\textit{Actor Net} and \textit{Critic Net} are two neural networks in PPO that take the intersection's information $\mathbf{S^{''}}$ as input. 
More specifically, the \textit{Actor Net} is used to predict the actions, that is the green duration for each phase, 
while the \textit{Critic Net} predicts the value for each state. 
To find the optimal policy, the PPO algorithm uses the ratio between the new and old policies scaled by the advantage to modify the objective function based on the actor-critic method \cite{konda1999actor, peters2008natural}. 
The PPO objective can be written as~\eqref{eq_menhod_PPO},

\begin{equation} \label{eq_menhod_PPO}
    J_{\theta} = \mathbb{E} \big[ \min(r_{t}(\theta)A_{t}, clip(r_{t}(\theta), 1-\varepsilon, 1+\varepsilon) A_{t}) \big],
\end{equation}

\noindent where $A_{t} = r_{t+1} + \gamma V(s_{t+1}) - V(s_{t})$ is denoted the advantage, 
and the $r_{t}(\theta) = \frac{\pi_{\theta}(a_{t}|s_{t})}{\pi_{\theta_{old}}(a_{t}|s_{t})}$ is the ratio between the new and the old policies. 
The clip function can prevent a large update to the policy weights.

The detailed network architecture in ADLight can be found in Table~\ref{tab_method_model_architectures}. 
For instance, ``$8 \times 1$ Conv $256$'' represents a convolutional layer that has 256 filters of size $8 \times 1$.

\begin{table}[!ht]
    \centering
    \caption{The network architecture of ADLight}
    \label{tab_method_model_architectures}
    \resizebox{0.7\textwidth}{!}{%
    \begin{tabular}{ccccc}
    \hline
    Module & Layer & Description & Input Size & Output Size \\ \hline
    \multirow{5}{*}{Feature Extractor} & Conv & $1 \times 8$ Conv 128, ReLU & $1 \times 8 \times 8$ & $128 \times 8 \times 1$ \\
     & Conv & $8 \times 1$ Conv 256, ReLU & $128 \times 8 \times 1$ & $256 \times 1 \times 1$ \\
     & Flatten & - & $256 \times 1 \times 1$ & $256$ \\
     & FC Layer & FC-128, ReLU & $256$ & $128$ \\
     & FC Layer & FC-64, ReLU & $128$ & $64$ \\ \hline
    \multirow{2}{*}{Policy Network} & FC Layer & FC-32, ReLU & $64$ & $32$ \\
     & FC Layer & FC-12 & $32$ & $12$ \\ \hline
    \multirow{2}{*}{Value Network} & FC Layer & FC-32, ReLU & $64$ & $32$ \\
     & FC Layer & FC-1 & $32$ & $1$ \\ \hline
    \end{tabular}%
    }
\end{table}

\section{Experiments} \label{experiments}

\subsection{Experiment Settings}

The experiments are conducted on the SUMO \cite{krajzewicz2002sumo}. 
SUMO is an open source, highly portable, microscopic, and continuous multi-modal traffic simulation platform \footnote{\url{https://sumo.dlr.de/docs/index.html}}. 
In this work, \textit{Traffic Control Interface} (TraCI) \footnote{\url{https://sumo.dlr.de/docs/TraCI.html}} is used to control traffic lights and retrieve information about traffic conditions for intersections. 
Induction loops detectors \footnote{\url{https://sumo.dlr.de/docs/Simulation/Output/Induction_Loops_Detectors_\%28E1\%29.html}} are used to retrieve traffic flow, 
and lane area detectors \footnote{\url{https://sumo.dlr.de/docs/Simulation/Output/Lanearea_Detectors_\%28E2\%29.html}} are used to detect lane occupancy and queue length. 
Noted that to suit the real-world scenario, the detector length is set to 100m instead of covering the entire lane. 
Also, a green light is followed by a 3s yellow light before it turns to a red light to ensure the safety of drivers.

\begin{figure}[!ht]
  \centering 
  \includegraphics[width=0.99\textwidth]{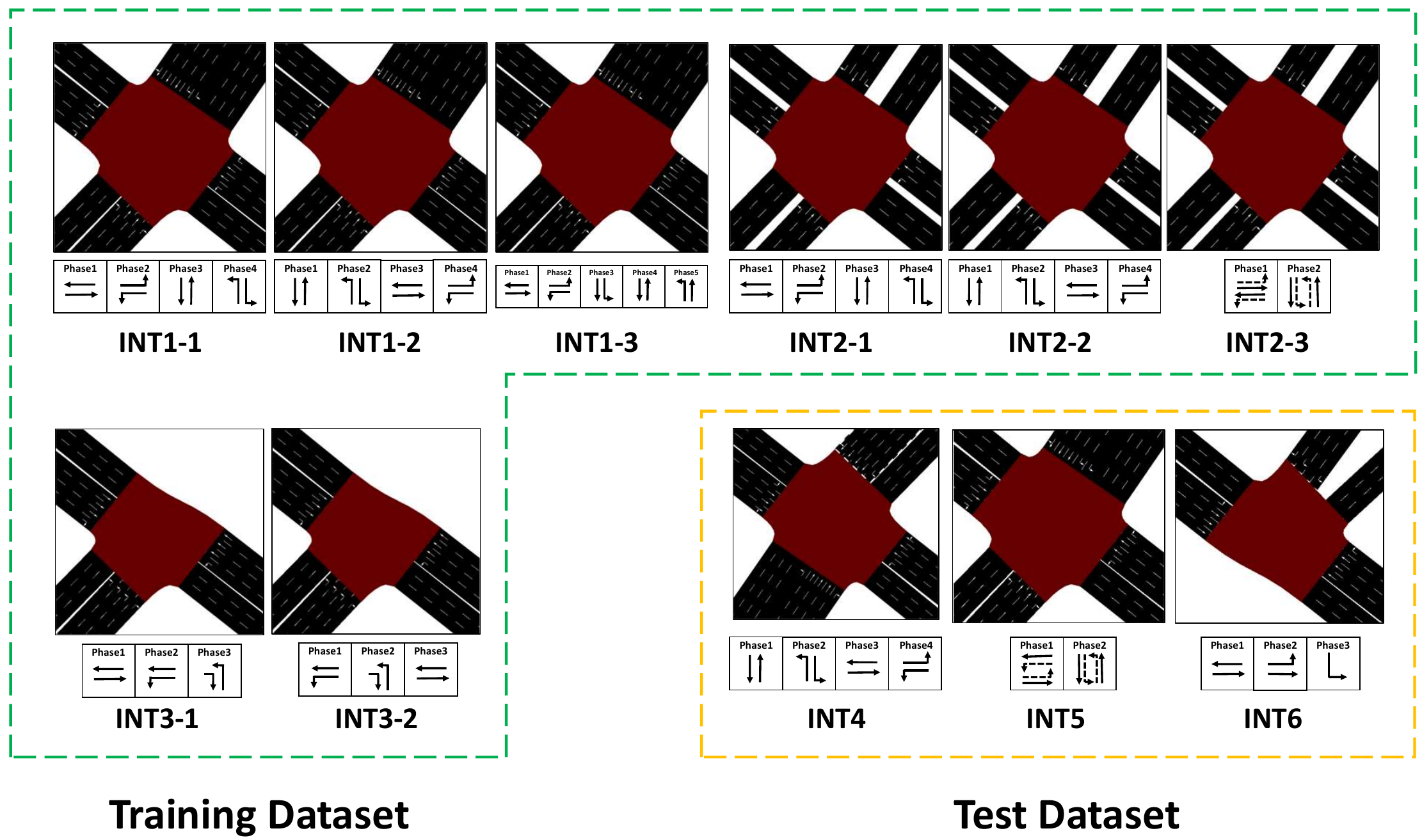}
  \caption{11 intersections with different topologies and phases in SUMO.} 
  \label{fig_experiment_dataset}
\end{figure}

\subsection{Dataset} \label{experiment_dataset}

This study considers intersections with different structures, aiming to employ one single universal model to predict actions for all intersections. 
Specifically, $11$ intersections of varying numbers of phases, 
lanes on each road, and approaching roads (i.e., 3-way or 4-way intersections) are constructed and used for experiments. 
Among these $11$ intersections, eight for training and three for testing. 
Figure~\ref{fig_experiment_dataset} shows the topologies and phases of the $11$ intersections 
while Table~\ref{tab_experiments_dataset} summarizes the configurations of all intersections. 

\begin{table}[!ht]
    \centering
    \caption{All intersection configurations.}
    \label{tab_experiments_dataset}
    \resizebox{0.99\textwidth}{!}{%
    \begin{tabular}{ccccccccclccc}
    \hline
     & \multicolumn{8}{c}{\textbf{Training Dataset}} &  & \multicolumn{3}{c}{\textbf{Test Dataset}} \\ \cline{2-9} \cline{11-13} 
    intersection ID & INT1-1 & INT1-2 & INT1-3 & INT2-1 & INT2-2 & INT2-3 & INT3-1 & INT3-2 &  & INT4 & INT5 & INT6 \\ \hline
    roads & 4 & 4 & 4 & 4 & 4 & 4 & 3 & 3 &  & 4 & 4 & 3 \\
    lanes per road & (5,4,4,4) & (5,4,4,4) & (5,4,4,4) & (3,3,3,3) & (3,3,3,3) & (3,3,3,3) & (0,4,4,4) & (0,4,4,4) &  & (5,4,5,4) & (5,4,4,4) & (4,4,4,0) \\
    phases & 4 & 4 & 5 & 4 & 4 & 2 & 3 & 3 &  & 4 & 2 & 3 \\ \hline
    \end{tabular}%
}
\end{table}

Taking INT1-1 as an example, as shown in Figure~\ref{fig_experiment_dataset}, 
it consists of four bi-directional approaching roads with five lanes in the north-south direction and four lanes in each of the other three directions. 

It includes four phases, each of which combines two different movement signals. 
Therefore, \textit{roads}, \textit{lanes per road} and \textit{phases} are $4$, $(5,4,4,4)$ clockwise and $4$ in Table~\ref{tab_experiments_dataset}, respectively.

There are three different intersection topologies in the training dataset, 
that is \textbf{INT1-X}, including INT1-1, INT1-2 and INT1-3, which represents a large 4-way intersection scenario, where each road has more than four lanes; 
\textbf{INT2-X}, including INT2-1, INT2-2 and INT2-3, represents a small 4-way intersection scenario, where each road has three lanes; 
\textbf{INT3-X}, including INT3-1 and INT2-2, represents the 3-way intersection scenario. 
For the intersections with the same topology, 
we change the sequences of the phases or the number of phases to generate new intersections. 
For example, the configurations for INT1-1 and INT1-2 are exactly the same, but the sequence of phases is different. 
For INT1-1 and INT1-3, the number of phases for these two intersections is different.

In order to validate the performance of the proposed model on unseen intersections, 
three testing scenarios are formed, namely INT4, INT5 and INT5. 
For instance, INT4 increases the number of lanes based on INT1 whereas  
INT5 changes the number of phases space from four to two on the basis of INT1. 
By rotating INT3, INT6 is obtained.

We also generate 100 unique pieces of route for each intersection, 
and three-quarters of the routes are used for training while the rest are for evaluation. 
The duration of each route is 30000s (about 8 hours).

\subsection{Evaluation Metric}

The waiting time per vehicle is used to evaluate the performance of different methods. 
The waiting time for each vehicle is defined as follow: 
the time duration when the vehicle’s speed is below or equal to 0.1m/s during the trip \footnote{\url{https://sumo.dlr.de/docs/Simulation/Output/Summary.html}}. 
A shorter waiting time means that vehicles spend less time passing through the intersection.

\subsection{Baseline Methods} \label{experiment_compared_method}

To evaluate the performance of the proposed ADLight, 
we compare the resulting universal model with the following existing methods, 
including both classical techniques in the transportation field and RL models with different action designs. 
For RL models, the agents are trained by PPO \cite{schulman2017proximal}, 
and the state and the reward design are the same as discussed in \nameref{methodology_agent_design}.

\textbf{Webster} \cite{koonce2008traffic}: 
Using traffic volume and phase sequence to calculate the cycle length and phase split. 
It can be proved that when the traffic is uniform, the Webster method minimizes the travel time of all vehicles passing the intersection or maximizes the intersection capacity \cite{weiSurveyTrafficSignal2020}. 
This Webster method can also be modified for real-time applications. 
For fairness, we use the Webster method to adjust the traffic lights based on real-time traffic in this study.

\textbf{Choose Next Phase (single-env)} \cite{oroojlooy2020attendlight, shabestaryAdaptiveTrafficSignal2022}: 
Using the action \textit{choose next phase} to train in a single environment. 
\textit{Choose next phase} means the agent can change to any legitimate phase in a variable phase sequence. 
This action design is flexible in that the traffic signal is non-cyclical. 
However, selecting phases randomly can confuse the drivers, potentially incurring traffic accidents.

\textbf{Next or Not (single-env)} \cite{van2016coordinated, weiIntelliLightReinforcementLearning2018}: 
Using the action \textit{next or not} to train in a single environment. 
When applying this action design, the agent decides to keep the current phase or change to the next phase. 
Compared with \textit{choose next phase}, \textit{next or not} can change the signal cyclically. 
However, the model trained with \textit{next or not} is related to the phase sequence of the intersection, 
which means the trained model cannot directly transfer from one intersection to another.

\textbf{Set Current Phase Duration (single-env)} \cite{xu2013study, aslaniTrafficSignalOptimization2018}: 
Using the action \textit{set current phase duration} to train in a single environment. 
\textit{Set current phase duration} means that the agent learns to set the duration for the current phase by choosing from pre-defined candidate time periods. 
Unlike the universal model trained by ADLight, this model is only trained on a single intersection, instead of multiple intersections.

\textbf{Set Current Phase Duration (multi-env)}: 
Using the action \textit{set current phase duration} to train in multiple environments. 
Although this method also trains on multiple intersections at the same time, compared to ADLight, it does not use the data augmentation shown in the second part of Figure~\ref{fig_methodology_framework}. 
This method is used to validate the effectiveness of the proposed data augmentation.

\subsection{Results of the training intersections with different routes}

By extracting features according to movement and using the \textit{set current phase duration} as the agent's action, our method has the same model structure on the intersection with different structures. 
Therefore, our universal model is trained on eight intersections before being tested on each intersection with different routes in this experiment. 
We compare ADLight with the Webster model, Choose Next Phase (single-env), 
Next or Not (single-env), Set Current Phase Duration (single-env) and Set Current Phase Duration (multi-env).

Table~\ref{tab_experiments_results_training_dataset_multi} shows the waiting time of ADLight and other universal models on the training intersections. 
It is clear that the RL-based approaches outperform the Webster approach. 
This is because the Webster model assumes that the traffic is uniform. 
Unfortunately, this assumption may not hold in reality. 
As a result, the calculated cycle and phase split will be inaccurate, thus leading to degraded performance.

When comparing the two universal methods, \textit{Set Current Phase Duration (multi-env)} and ADLight, both of which are trained in multiple environments, the performance of ADLight is improved by an average of $10\%$ on eight intersections, as compared to \textit{Set Current Phase Duration (multi-env)}. 
This confirms that the proposed new data augmentation method can improve the performance of universal models.

\begin{table}[!ht]
    \centering
    \caption{Results of training intersections for universal models}
    \label{tab_experiments_results_training_dataset_multi}
    \resizebox{0.99\textwidth}{!}{%
    \begin{tabular}{ccccccccc}
    \hline
    Method Name & INT1-1 & INT1-2 & INT1-3 & INT2-1 & INT2-2 & INT2-3 & INT3-1 & INT3-2 \\ \hline
    Webster & 55.1955 & 53.8775 & 59.9905 & 47.246 & 47.3615 & 10.391 & 9.7525 & 9.825 \\
    Set Current Phase Duration (multi-env) & 10.8135 & 9.443 & 12.9 & 12.895 & 12.9885 & 5.274 & 3.831 & 3.8415 \\
    \textbf{ADLight} & 9.9505 & 8.8435 & 11.562 & 11.738 & 11.7275 & 3.9385 & 3.8855 & 3.888 \\ \hline
    \end{tabular}%
    }
\end{table}

Table~\ref{tab_experiments_results_training_dataset_single} shows the performance of RL models trained on a specific intersection instance using three different action designs. 
It can be seen that the action \textit{choose next phase} can achieve the minimum waiting time per vehicle in most cases since it can choose any phase arbitrarily every $N$ seconds ($N=5$ in this study). 
However, this disordered phase combination incurs safety concerns for drivers as mentioned in Section~\nameref{experiment_compared_method}. 
When dealing with the intersection of only two phases, 
\textit{choose next phase} will degenerate to other actions. 
For example, in intersection INT2-3, the performance of \textit{choose next phase} is not as good as other actions. 
At the same time, the result of \textit{set current phase duration} is better than \textit{next or not} in most cases. 

\begin{table}[!ht]
    \centering
    \caption{Results of training intersections for single-env models}
    \label{tab_experiments_results_training_dataset_single}
    \resizebox{0.99\textwidth}{!}{%
    \begin{tabular}{ccccccccc}
    \hline
    Method Name & INT1-1 & INT1-2 & INT1-3 & INT2-1 & INT2-2 & INT2-3 & INT3-1 & INT3-2 \\ \hline
    Choose Next Phase (single-env) & 6.757 & 6.689 & 6.312 & 6.2115 & 6.147 & 1.734 & 2.815 & 2.955 \\
    Next or Not (single-env) & 10.0645 & 10.2045 & 11.5015 & 10.384 & 10.476 & 1.707 & 3.7745 & 3.835 \\
    Set Current Phase Duration (single-env) & 9.0315 & 7.9295 & 10.29 & 9.6085 & 9.4635 & 1.289 & 3.096 & 3.1045 \\ \hline
    \end{tabular}%
    }
\end{table}

Finally, we compare the performance of the models trained in a single environment (as shown in Table~\ref{tab_experiments_results_training_dataset_single}) with those trained in multiple environments  (as shown in Table~\ref{tab_experiments_results_training_dataset_multi}). 
The comparison of Table 3 and Table 4 shows that the multi-env modes are more versatile at the cost of marginal performance degradation in terms of the waiting time of about 2.3s as compared to the single-env models. 
This performance loss is acceptable as the universal models take into account consider all intersections. 
It should be emphasized that the multi-env models only need to be trained once before being tested on eight different intersections without further training.  
In contrast, the single-env models have to be trained on each intersection. As a result, the training time for the single-env models is eight times longer than that of the multi-env models. 
When thousands of intersections are considered, the training time for single-env models will be even more problematic.

\subsection{Results of the test intersections}

We evaluate the key feature of ADLight to see whether it can be utilized for new unseen intersections directly. 
Three intersections, INT4, INT5 and INT6, are used for testing. 
As discussed in Section~\nameref{experiment_dataset}, these intersections differ from the scenarios in the training set in terms of the number of lanes or the number of phases. 
To the best of our knowledge, no RL models using \textit{set current phase duration} in the literature can work on more than one intersection scenarios without any retraining or transfer learning. 
For comparison purposes, we compare the proposed ADLight against the model trained on the three testing intersections. 
At the same time, by applying the agent design discussed in Section~\nameref{methodology_agent_design}, a model with the same structure can suit the intersections with different structures. 
Therefore, the models trained on a single scenario on the training set can also be compared with ADLight to explore whether the inclusion of diverse intersections in the training set would help model performance.

\begin{table}[!ht]
    \centering
    \caption{Result of test intersections}
    \label{tab_experiments_results_test_dataset}
    \resizebox{0.95\textwidth}{!}{%
    \begin{tabular}{ccccc}
    \hline
    Method Name & INT4 & INT5 & INT6 & \textbf{Degradation} \\ \hline
    Webster & 156.5445 & 11.045 & 10.5905 &  \\ \hline
    \multicolumn{1}{l}{} & \multicolumn{4}{c}{\textbf{Models train in the single-env and test in the same environments}} \\
    Set Current Phase Duration (single-env) & 13.1445 & 1.8475 & 3.8045 &  \\ \hline
    \multicolumn{1}{l}{} & \multicolumn{4}{c}{\textbf{Models train in the single-env and test in other environments}} \\
    Set Current Phase Duration (INT1-1) & 15.475 ($\uparrow 17.73\%$) & 2.881 ($\uparrow 55.94\%$) & 5.68 ($\uparrow 49.297\%$) & $\uparrow 40.989\%$ \\
    Set Current Phase Duration (INT2-3) & 17.858 ($\uparrow 35.855\%$) & 1.846 ($\downarrow 0.108\%$) & 6.285 ($\uparrow 65.199\%$) & $\uparrow 33.649\%$ \\
    Set Current Phase Duration (INT3-1) & 21.514 ($\uparrow 63.673\%$) & 2.034 ($\uparrow 10.095\%$) & 5.364 ($\uparrow 40.991\%$) & $\uparrow 38.253\%$ \\ \hline
    \multicolumn{1}{l}{} & \multicolumn{4}{c}{\textbf{Models train in multi-env}} \\
    Set Current Phase Duration (multi-env) & 19.3945 ($\uparrow 47.548\%$) & 3.2125 ($\uparrow 73.884\%$) & 6.427 ($\uparrow 68.932\%$) & $\uparrow 63.455\%$ \\
    \textbf{ADLight} & 16.917 ($\uparrow 28.7\%$) & 1.944 ($\uparrow 5.223\%$) & 5.299 ($\uparrow 39.282\%$) & $\uparrow 24.402\%$ \\ \hline
    \end{tabular}%
    }
\end{table}

Table~\ref{tab_experiments_results_test_dataset} shows the performance of different models on the test dataset. 
We also calculate the performance between different methods with respect to the single-env model.  
The degradation represents the increase in waiting time over the single-environment model. 
Similar to the results on the training set, 
the performance of the RL-based algorithms is still far superior to traditional traffic control algorithms. 
When testing on the unseen environment instances, 
the model degradation between the multi-env models and the single-env models is $63\%$. 
After using \textit{movement shuffle}, the performance of the proposed method ADLight improves, reducing the waiting time loss to $24\%$ on average.

Next, we analyze whether including more intersections in the training set can help the model improve its performance. 
As shown in Table~\ref{tab_experiments_results_test_dataset}, simply training in multiple environments incurs performance loss, which can be effectively alleviated by adding data augmentation. 
At the same time, the models trained in the single environment can also be transferred to new intersections. 
In particular, the large similarity between the unseen scene and the training scene leads to less performance degradation. 
For example, the model trained on INT1-1 can perform well in INT4 compared with INT5 and INT6, as both INT1-1 and INT4 are 4-way intersections with the same phase structure. 
The only difference between INT1-1 and INT4 is the number of lanes. 
In contrast, changing the number of phases (e.g., from 4 phases in INT1-1 to 2 phases in INT5) or the number of roads (e.g., from the 4-way intersection in INT1-1 to the 3-way intersection in INT6) incurs significant drop in performance.

\subsection{Results of retraining in test intersections}

In practice, there are some important intersections that we need to pay more attention to their performance. 
For this purpose, we start from the universal model trained by ADLight. 
Compared to the single-env model, the resulting model can obtain a quite small performance degradation in an unseen environment only after a few training steps.

\begin{figure}[ht!]
    \centering
    \subfloat[INT4]{
        \includegraphics[width=0.45\textwidth]{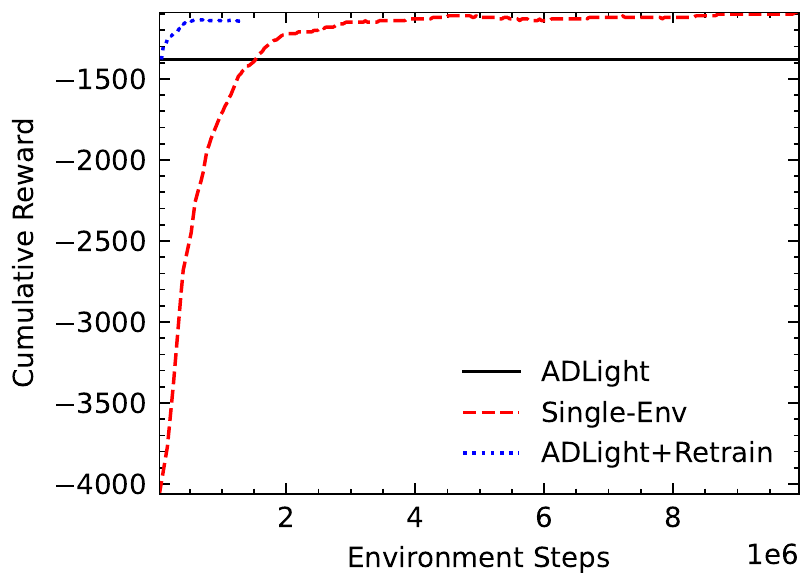}
        \label{fig_exoeriment_INT4}
    }
    \subfloat[INT5]{
        \includegraphics[width=0.45\textwidth]{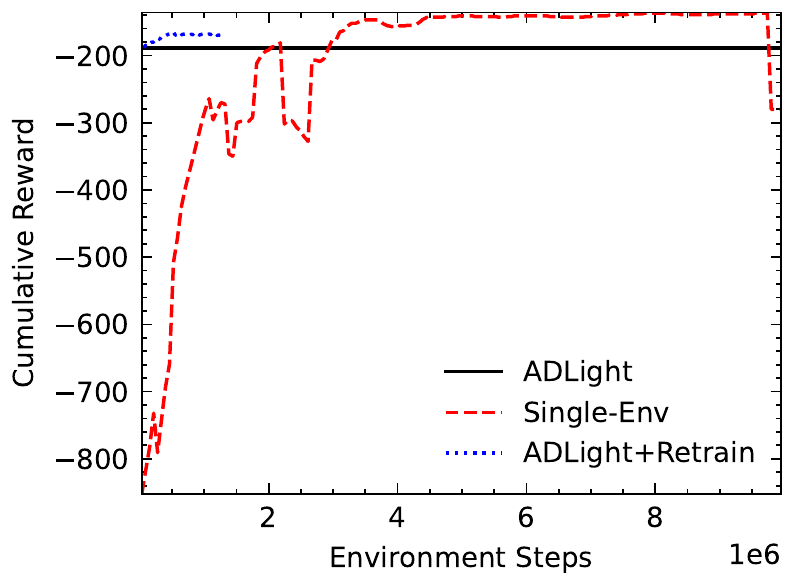}
        \label{fig_exoeriment_INT5}
    }
    \quad
    \subfloat[INT6]{
        \includegraphics[width=0.47\textwidth]{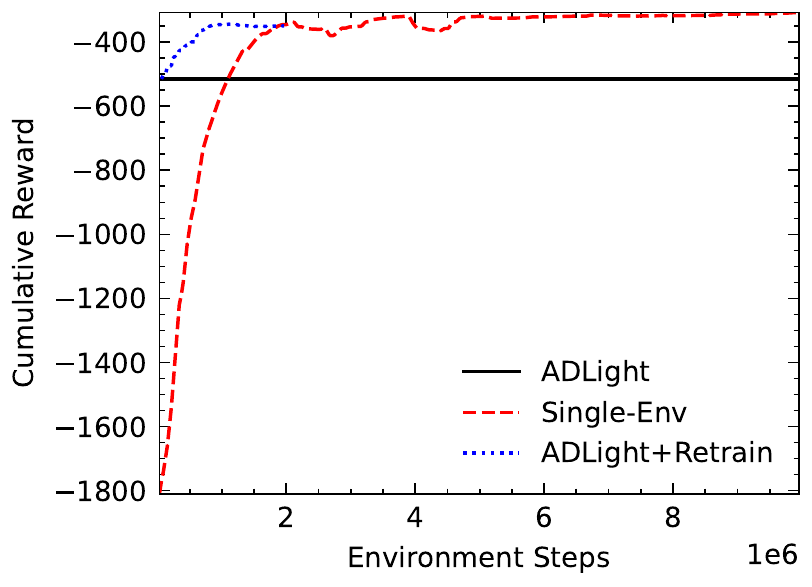}
        \label{fig_exoeriment_INT6}
    }
    \caption{The environment steps of different models at the three test intersections.} 
    \label{fig_training_steps}
\end{figure}

Figure~\ref{fig_training_steps} shows the training steps and the corresponding cumulative rewards for different models in the test intersections. 
The black line demonstrates the result of applying the generalized model directly at the new intersections without retraining. 
It can be seen that the model already demonstrates good results without any transfer learning. 
The red dashed line represents the model trained from scratch, while the blue dashed line represents the model trained based on the universal model. 
At about $6 \times 10^{6}$ training steps, the single-env model converges while the retrained model only requires around $10^{6}$ training steps, 
which amounts to about an $80\%$ saving of the computation time while maintaining comparable performance. 

\begin{table}[!ht]
    \centering
    \caption{Results of retraining in test intersections}
    \label{tab_experiments_results_retraining}
    \resizebox{0.95\textwidth}{!}{%
    \begin{tabular}{ccccc}
    \hline
    Method Name & INT4 & INT5 & INT6 & \textbf{Degradation} \\ \hline
    Set Current Phase Duration (single-env) & 13.1445 & 1.8475 & 3.8045 &  \\
    Set Current Phase Duration (multi-env) & 19.3945 ($\uparrow 47.548\%$) & 3.2125 ($\uparrow 73.884\%$) & 6.427 ($\uparrow 68.932\%$) & $\uparrow 63.455\%$ \\
    ADLight & 16.917 ($\uparrow 28.7\%$) & 1.944 ($\uparrow 5.223\%$) & 5.299 ($\uparrow 39.282\%$) & $\uparrow 24.402\%$ \\
    \textbf{ADLight + Retrain} & 13.5725 ($\uparrow 3.256\%$) & 1.939 ($\uparrow 4.953\%$) & 4.1055 ($\uparrow 7.912\%$) & $\uparrow 5.373\%$ \\ \hline
    \end{tabular}
    }
\end{table}

Table~\ref{tab_experiments_results_retraining} shows the detailed model performance after retraining. 
After $10^{6}$ training steps, the performance degradation drops from $24\%$ to $5\%$, 
while training a model from scratch needs to take about at least $6 \times 10^{6}$ training steps. 
This is particularly attractive for real-time applications. 
For example, in a road network of more than 1000 junctions, at least $5 \times 10^{9}$ interactions with environments can be reduced, which greatly improves the training efficiency.

\section{Conclusion} \label{conclusion}

This paper proposes an approach called ADLight that trains a universal model under a comprehensive set of intersections and transfers to unseen intersections. 
In other words, once the model is trained, it can be used for new intersections and provide reasonable performance. 
In ADLight, the agent's state and action are specially designed to ensure that the same model structure can handle different intersections. 
A new data augmentation method named \textit{movement shuffle} is designed for the signal light control problem. 
The experiments show that by using this data augmentation method 
the performance degradation decreases from $63\%$ to $24\%$ as compared with single-env models in unseen intersections. 
In addition, the performance degradation can drop from $24\%$ to $5\%$ only after a few steps of retraining on the universal model, this reduces nearly $80\%$ of training time as compared to training from scratch.

This work can be extended in several directions. 
First, more intersections with different structures can be added to enhance the generality of the model. 
Second, waiting time per vehicle should not be the only metric to evaluate the performance of different methods. 
Metrics such as robustness, safety, and comfort should be also taken into consideration. 
Finally, ADLight can be further extended to control multiple intersections in the same area.

\section{Acknowledgements}

This work was supported in part by the Shenzhen Science and Technology Innovation Committee under Grant No. JCYJ20190813170803617 and the Shanghai Pujiang Program under Grant No. 21PJD092.

\bibliographystyle{unsrt}  
\bibliography{references}

\begin{thebibliography}{10}

\bibitem{kweku2018greenhouse}
Darkwah~Williams Kweku, Odum Bismark, Addae Maxwell, Koomson~Ato Desmond,
  Kwakye~Benjamin Danso, Ewurabena~Asante Oti-Mensah, Asenso~Theophilus
  Quachie, and Buanya~Beryl Adormaa.
\newblock Greenhouse effect: greenhouse gases and their impact on global
  warming.
\newblock {\em Journal of Scientific research and reports}, 17(6):1--9, 2018.

\bibitem{miller1963settings}
Alan~J Miller.
\newblock Settings for fixed-cycle traffic signals.
\newblock {\em Journal of the Operational Research Society}, 14(4):373--386,
  1963.

\bibitem{koonce2008traffic}
Peter Koonce and Lee Rodegerdts.
\newblock Traffic signal timing manual.
\newblock Technical report, United States. Federal Highway Administration,
  2008.

\bibitem{gershenson2004self}
Carlos Gershenson.
\newblock Self-organizing traffic lights.
\newblock {\em arXiv preprint nlin/0411066}, 2004.

\bibitem{roess2004traffic}
Roger~P Roess, Elena~S Prassas, and William~R McShane.
\newblock {\em Traffic engineering}.
\newblock Pearson/Prentice Hall, 2004.

\bibitem{shabestaryAdaptiveTrafficSignal2022}
Soheil Mohamad~Alizadeh Shabestary and Baher Abdulhai.
\newblock Adaptive {Traffic} {Signal} {Control} {With} {Deep} {Reinforcement}
  {Learning} and {High} {Dimensional} {Sensory} {Inputs}: {Case} {Study} and
  {Comprehensive} {Sensitivity} {Analyses}.
\newblock {\em IEEE Transactions on Intelligent Transportation Systems}, pages
  1--15, 2022.
\newblock ZSCC: 0000000.

\bibitem{xu2013study}
Lun-Hui Xu, Xin-Hai Xia, and Qiang Luo.
\newblock The study of reinforcement learning for traffic self-adaptive control
  under multiagent markov game environment.
\newblock {\em Mathematical Problems in Engineering}, 2013, 2013.

\bibitem{mannion2016experimental}
Patrick Mannion, Jim Duggan, and Enda Howley.
\newblock An experimental review of reinforcement learning algorithms for
  adaptive traffic signal control.
\newblock {\em Autonomic road transport support systems}, pages 47--66, 2016.

\bibitem{van2016coordinated}
Elise Van~der Pol and Frans~A Oliehoek.
\newblock Coordinated deep reinforcement learners for traffic light control.
\newblock {\em Proceedings of learning, inference and control of multi-agent
  systems (at NIPS 2016)}, 1, 2016.

\bibitem{weiIntelliLightReinforcementLearning2018}
Hua Wei, Guanjie Zheng, Huaxiu Yao, and Zhenhui Li.
\newblock {IntelliLight}: {A} {Reinforcement} {Learning} {Approach} for
  {Intelligent} {Traffic} {Light} {Control}.
\newblock In {\em Proceedings of the 24th {ACM} {SIGKDD} {International}
  {Conference} on {Knowledge} {Discovery} \& {Data} {Mining}}, pages
  2496--2505, London United Kingdom, July 2018. ACM.
\newblock 00000.

\bibitem{weiPressLightLearningMax2019}
Hua Wei, Chacha Chen, Guanjie Zheng, Kan Wu, Vikash Gayah, Kai Xu, and Zhenhui
  Li.
\newblock {PressLight}: {Learning} {Max} {Pressure} {Control} to {Coordinate}
  {Traffic} {Signals} in {Arterial} {Network}.
\newblock In {\em Proceedings of the 25th {ACM} {SIGKDD} {International}
  {Conference} on {Knowledge} {Discovery} \& {Data} {Mining}}, pages
  1290--1298, Anchorage AK USA, July 2019. ACM.
\newblock ZSCC: 0000097.

\bibitem{weiCoLightLearningNetworklevel2019}
Hua Wei, Nan Xu, Huichu Zhang, Guanjie Zheng, Xinshi Zang, Chacha Chen, Weinan
  Zhang, Yanmin Zhu, Kai Xu, and Zhenhui Li.
\newblock {CoLight}: learning network-level cooperation for traffic signal
  control.
\newblock In {\em Proceedings of the 28th {ACM} {International} {Conference} on
  {Information} and {Knowledge} {Management}}, pages 1913--1922, Beijing China,
  November 2019. ACM.
\newblock 00000.

\bibitem{zhengLearningPhaseCompetition2019}
Guanjie Zheng, Yuanhao Xiong, Xinshi Zang, Jie Feng, Hua Wei, Huichu Zhang,
  Yong Li, Kai Xu, and Zhenhui Li.
\newblock Learning {Phase} {Competition} for {Traffic} {Signal} {Control}.
\newblock In {\em Proceedings of the 28th {ACM} {International} {Conference} on
  {Information} and {Knowledge} {Management}}, pages 1963--1972, Beijing China,
  November 2019. ACM.
\newblock ZSCC: 0000066.

\bibitem{liangDeepReinforcementLearning2019}
Xiaoyuan Liang, Xunsheng Du, Guiling Wang, and Zhu Han.
\newblock A {Deep} {Reinforcement} {Learning} {Network} for {Traffic} {Light}
  {Cycle} {Control}.
\newblock {\em IEEE Transactions on Vehicular Technology}, 68(2):1243--1253,
  February 2019.

\bibitem{chenThousandLightsDecentralized2020}
Chacha Chen, Hua Wei, Nan Xu, Guanjie Zheng, Ming Yang, Yuanhao Xiong, Kai Xu,
  and Zhenhui Li.
\newblock Toward {A} {Thousand} {Lights}: {Decentralized} {Deep}
  {Reinforcement} {Learning} for {Large}-{Scale} {Traffic} {Signal} {Control}.
\newblock {\em Proceedings of the AAAI Conference on Artificial Intelligence},
  34(04):3414--3421, April 2020.
\newblock 00000.

\bibitem{ma2021adaptive}
Zian Ma, Chengcheng Xu, Yuheng Kan, Maonan Wang, and Wei Wu.
\newblock Adaptive coordinated traffic control for arterial intersections based
  on reinforcement learning.
\newblock In {\em 2021 IEEE International Intelligent Transportation Systems
  Conference (ITSC)}, pages 2562--2567. IEEE, 2021.

\bibitem{mousavi2017traffic}
Seyed~Sajad Mousavi, Michael Schukat, and Enda Howley.
\newblock Traffic light control using deep policy-gradient and
  value-function-based reinforcement learning.
\newblock {\em IET Intelligent Transport Systems}, 11(7):417--423, 2017.

\bibitem{aslani2017adaptive}
Mohammad Aslani, Mohammad~Saadi Mesgari, and Marco Wiering.
\newblock Adaptive traffic signal control with actor-critic methods in a
  real-world traffic network with different traffic disruption events.
\newblock {\em Transportation Research Part C: Emerging Technologies},
  85:732--752, 2017.

\bibitem{aslaniTrafficSignalOptimization2018}
Mohammad Aslani, Stefan Seipel, Mohammad~Saadi Mesgari, and Marco Wiering.
\newblock Traffic signal optimization through discrete and continuous
  reinforcement learning with robustness analysis in downtown {Tehran}.
\newblock {\em Advanced Engineering Informatics}, 38:639--655, October 2018.
\newblock 00000.

\bibitem{xiongLearningTrafficSignal2019}
Yuanhao Xiong, Guanjie Zheng, Kai Xu, and Zhenhui Li.
\newblock Learning {Traffic} {Signal} {Control} from {Demonstrations}.
\newblock In {\em Proceedings of the 28th {ACM} {International} {Conference} on
  {Information} and {Knowledge} {Management}}, pages 2289--2292, Beijing China,
  November 2019. ACM.

\bibitem{yangCooperativeTrafficSignal2019}
Shantian Yang, Bo~Yang, Hau-San Wong, and Zhongfeng Kang.
\newblock Cooperative traffic signal control using {Multi}-step return and
  {Off}-policy {Asynchronous} {Advantage} {Actor}-{Critic} {Graph} algorithm.
\newblock {\em Knowledge-Based Systems}, 183:104855, November 2019.
\newblock ZSCC: 0000031.

\bibitem{oroojlooy2020attendlight}
Afshin Oroojlooy, Mohammadreza Nazari, Davood Hajinezhad, and Jorge Silva.
\newblock Attendlight: Universal attention-based reinforcement learning model
  for traffic signal control.
\newblock {\em Advances in Neural Information Processing Systems},
  33:4079--4090, 2020.

\bibitem{krajzewicz2002sumo}
Daniel Krajzewicz, Georg Hertkorn, Christian R{\"o}ssel, and Peter Wagner.
\newblock Sumo (simulation of urban mobility)-an open-source traffic
  simulation.
\newblock In {\em Proceedings of the 4th middle East Symposium on Simulation
  and Modelling (MESM20002)}, pages 183--187, 2002.

\bibitem{weiSurveyTrafficSignal2020}
Hua Wei, Guanjie Zheng, Vikash Gayah, and Zhenhui Li.
\newblock A {Survey} on {Traffic} {Signal} {Control} {Methods}, January 2020.

\bibitem{lowrie1990scats}
P~Lowrie.
\newblock Scats-a traffic responsive method of controlling urban traffic.
\newblock {\em Sales information brochure published by Roads \& Traffic
  Authority, Sydney, Australia}, 1990.

\bibitem{mnih2013playing}
Volodymyr Mnih, Koray Kavukcuoglu, David Silver, Alex Graves, Ioannis
  Antonoglou, Daan Wierstra, and Martin Riedmiller.
\newblock Playing atari with deep reinforcement learning.
\newblock {\em arXiv preprint arXiv:1312.5602}, 2013.

\bibitem{wang2020deep}
Hao-nan Wang, Ning Liu, Yi-yun Zhang, Da-wei Feng, Feng Huang, Dong-sheng Li,
  and Yi-ming Zhang.
\newblock Deep reinforcement learning: a survey.
\newblock {\em Frontiers of Information Technology \& Electronic Engineering},
  21(12):1726--1744, 2020.

\bibitem{haydariDeepReinforcementLearning2020}
Ammar Haydari and Yasin Yilmaz.
\newblock Deep {Reinforcement} {Learning} for {Intelligent} {Transportation}
  {Systems}: {A} {Survey}.
\newblock {\em IEEE Transactions on Intelligent Transportation Systems}, pages
  1--22, 2020.
\newblock ZSCC: 0000114.

\bibitem{cobbe2019quantifying}
Karl Cobbe, Oleg Klimov, Chris Hesse, Taehoon Kim, and John Schulman.
\newblock Quantifying generalization in reinforcement learning.
\newblock In {\em International Conference on Machine Learning}, pages
  1282--1289. PMLR, 2019.

\bibitem{lee2019network}
Kimin Lee, Kibok Lee, Jinwoo Shin, and Honglak Lee.
\newblock Network randomization: A simple technique for generalization in deep
  reinforcement learning.
\newblock {\em arXiv preprint arXiv:1910.05396}, 2019.

\bibitem{schulman2017proximal}
John Schulman, Filip Wolski, Prafulla Dhariwal, Alec Radford, and Oleg Klimov.
\newblock Proximal policy optimization algorithms.
\newblock {\em arXiv preprint arXiv:1707.06347}, 2017.

\bibitem{konda1999actor}
Vijay Konda and John Tsitsiklis.
\newblock Actor-critic algorithms.
\newblock {\em Advances in neural information processing systems}, 12, 1999.

\bibitem{peters2008natural}
Jan Peters and Stefan Schaal.
\newblock Natural actor-critic.
\newblock {\em Neurocomputing}, 71(7-9):1180--1190, 2008.

\end{thebibliography}

\end{document}